# SWIPE: a bolometric polarimeter for the Large-Scale Polarization Explorer

P. de Bernardis*[a], S. Aiola[a], G. Amico[a], E. Battistelli[a], A. Coppolecchia[a], A. Cruciani[a],
A. D' Addabbo[a], G. D' Alessandro[a], S. De Gregori[a], M. De Petris[a], D. Goldie[b], R. Gualtieri[a],
V. Haynes[c], L. Lamagna[a], B. Maffei[c], S. Masi[a], F. Nati[a], M. Wah Ng[c], L. Pagano[a], F. Piacentini[a], L. Piccirillo[c], G. Pisano[c], G. Romeo[d], M. Salatino[a], A. Schillaci[a], E. Tommasi[e], S. Withington[b]
[a]Dipartimento di Fisica, Sapienza Università di Roma, P.le A. Moro 2, 00185 Roma, Italy
[b]Cavendish Laboratory, University of Cambridge, JJ Thomson Avenue, Cambridge CB3 0HE, UK
[c]Jodrell Bank Centre for Astrophysics, University of Manchester, Macclesfield, SK11 9DL, UK
[d]Istituto Nazionale di Geofisica e Vulcanologia, Via di Vigna Murata, 605, 00143 Roma, Italy
[e]Agenzia Spaziale Italiana, Viale Liegi 26, 00198 Roma, Italy

Abstract

The balloon-borne LSPE mission is optimized to measure the linear polarization of the Cosmic Microwave Background at large angular scales. The Short Wavelength Instrument for the Polarization Explorer (SWIPE) is composed of 3 arrays of multi-mode bolometers cooled at 0.3K , with optical components and filters cryogenically cooled below 4K to reduce the background on the detectors. Polarimetry is achieved by means of large rotating half-wave plates and wire-grid polarizers in front of the arrays. The polarization modulator is the first component of the optical chain, reducing significantly the effect of instrumental polarization. In SWIPE we trade angular resolution for sensitivity. The diameter of the entrance pupil of the refractive telescope is 45 cm, while the field optics is optimized to collect tens of modes for each detector, thus boosting the absorbed power. This approach results in a FWHM resolution of 1.8, 1.5, 1.2 degrees at 95, 145, 245 GHz respectively. The expected performance of the three channels is limited by photon noise, resulting in a final sensitivity around 0.1-0.2 μK per beam, for a 13 days survey covering 25% of the sky.

## 1. INTRODUCTION

The Large Scale Polarization Explorer balloon-borne mission aims at measuring CMB polarization at large angular scales, with high sensitivity and wide frequency coverage. The SWIPE instrument on board of LSPE covers the frequency region 80 – 290 GHz , using cryogenic bolometers as detectors. At variance with other bolometric balloon-borne missions, like EBEX[1] and SPIDER[2], which use a very large number of diffraction-limited detectors, SWIPE uses a limited number of bolometers, which are multi-mode to achieve comparable sensitivity. This approach simplifies the detector array and its readout, avoiding multiplexing, at the cost of a coarser angular resolution (which is of the order of 2º FWHM). In this paper we describe the scientific targets of the SWIPE instrument, and its design.

## 2. THE SCIENCE CASE FOR SWIPE

Accurate measurements of CMB polarization at large angular scales are potentially sensitive to the rotational component (B-modes) produced by the inflation process, and could provide crucial information, like the energy scale of the inflation process (see e.g.[3] and references therein). Most of this information is encoded in the first 20-30 multipoles of the spherical harmonics expansion of the CMB polarization field: for this reason large angular-scale polarization measurements are of special interest (see the companion paper[4] and references therein). However, the B-modes signal is very small compared to both instrumental noise and polarized foregrounds. In fact, measurements of CMB polarization are heavily affected by polarized foregrounds from the interstellar medium in our Galaxy and in other galaxies. At frequencies below 60 GHz is important the effect of synchrotron emission, while above 90 GHz polarized interstellar emission (ISD) is the main contaminant. ISD is composed of populations of asymmetric spinning grains, whose largest inertia axis is dynamically aligned to the interstellar magnetic field lines. This results in preferential emission orthogonal to the magnetic field direction. The resulting degree of linearly polarized emission from ISD is of the order of a few % ([5,6]) as measured e.g. by Archeops[7] and WMAP[8]. In the best sky areas, anisotropic emission of the ISD is a few % of CMB anisotropy at 145 GHz (see e.g.[9]), so the expected polarized signal from ISD is of the same order of magnitude of the E-modes in CMB polarization. In this situation, an instrument aimed at measuring B-modes at large angular scales should be optimized to achieve high sensitivity while covering a wide spectral interval. In this way, CMB polarization can be accurately measured and separated from ISD, taking advantage of their different spectral dependence (see e.g. [10,11,12] and references therein). The SWIPE instrument on LSPE has been optimized to cover with very high sensitivity the high frequency side of the CMB spectrum. In this range ISD and CMB polarization produce comparable signals, and our target is to separate cleanly the two components. Combining STRIP[13], which covers the low-frequency side of the CMB spectrum with orthogonal technologies, LSPE becomes a powerful mission, with unprecedented sensitivity and efficient control of systematic effects.

## 3. SWIPE INSTRUMENT DESIGN

SWIPE is a sky-scanning Stokes polarimeter, using a rotating half-wave plate (HWP) and a steady polarizer in front of the detectors to modulate the linear polarization component of the incoming brightness. The presence of a polarization modulator is especially important for a scanning instrument measuring sky brightness at the largest angular scales. This approach has already been used by MAXIPOL[14] and is planned for the forthcoming EBEX[1], SPIDER[2], POLARBEAR[15], QUBIC[16]. Other bolometric polarimeters for the CMB compare the power measured by two detectors sensitive to the two orthogonal polarizations (see e.g. [7,17,18]). In such a configuration, however, uncorrelated drifts in detectors response and offset convert directly into apparent sky polarization. This systematic effect is mitigated by the use of matched couples of detectors[19]. In a Stokes polarimeter, instead, the same detector is used to measure both orthogonal polarizations, thus reducing significantly the systematic effects produced by instrumental drifts. Since drifts lie intrinsically at low frequencies in the time-ordered data, and in a scanning telescope the largest scales are observed at the lowest frequencies, the Stokes polarimeter approach is preferable. The observation strategy of a scanning Stokes polarimeter is described by two parameters: the sky-scan speed and the rotation rate of the HWP. The optimization of these two parameters has been discussed e.g. in[20]. In LSPE, the sky-scan speed is around 18°/s, as a tradeoff between the requirements of STRIP and SWIPE. At this scan speed, each independent pixel of the sky is observed for a fraction of a second. To achieve a significant number of polarization modulations in a fraction of a second the HWP should rotate at a rate of many revolutions per second, and should be cryogenically cooled to mitigate effects related to its emission. Fast rotating waveplates have been built using magnetically levitating devices[21]. In our system, instead, we simplify the modulator using a step and integrate HWP rotation approach: the HWP is rotated by a given angular step after a number of full revolutions of the gondola (say every 3 minutes, i.e. after 9 revolutions), and kept steady otherwise. After acquiring scans with HWP angles of say 0°, 22°.5, 45° and 67°.5, enough data are accumulated to retrieve the linear polarization information for each of the pixels of the scanned region. For the four different orientations of the waveplate, the scan of a slightly polarized sky produces a slightly different shape of the signal detected along the scan.

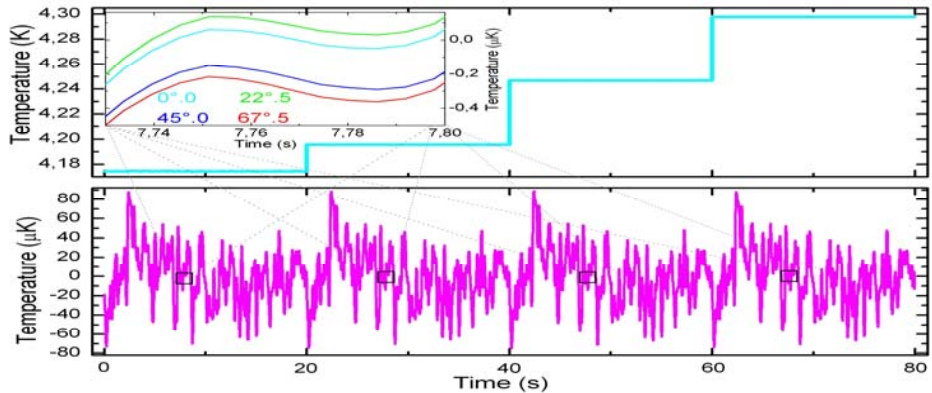

Figure 1. **Top:** Noiseless simulation of the measurement of microwave emission from the CMB, the instrument, and the stratosphere, at 145 GHz (in CMB temperature units) during 4 revolutions of the payload. In this simulation, at the end of each revolution the HWP is stepped by 22°.5. Non-idealities of the HWP and its thermal emission produce large steps in the detected power. **Bottom:** the same data once the averages over each revolution and the dipole have been removed. CMB anisotropy is evident. **Inset:** detail of the signal detected from the same sample sky region for 4 HWP orientations.

Systematic effects related to non-idealities in the HWP - which can be quite important[22,23] - produce variations of the general offset of each scan, which is easily removed from the time-ordered data (see Fig.1). On the other hand, the modulation of the polarized component of the sky in our baseline strategy has a cycle 12 minutes long, so a very stable system is required. The performance of this approach has been investigated using numerical simulations, including instrument drifts and 1/f noise, and confirming its effectiveness in removing the systematic effects due to non-idealities and emission of the HWP. High angular resolution is not a priority for this experiment, aimed at measuring CMB polarization at large angular scales. However, our target is to have the cutoff in the beam transfer function at multipoles larger than $\ell$=100, so that SWIPE can be calibrated against Planck data at the same frequency and at the same angular scales. For this reason, we target an angular resolution of about $1.5^\circ$-$2^\circ$ FWHM, so that the beam transfer function for multipole $\ell$=100 is still around 35-50%, while for multipole $\ell$=20 it is more than 95% (assuming Gaussian beams). In the case of non-gaussian, flatter-top beams, as expected for multi-mode optics, the values of the beam transfer function will be even larger. For a given angular resolution, the number of modes we want to collect on each detector sets the diameter of the entrance aperture of the polarimeter. With an aperture diameter of about 500 mm (the maximum reasonable size for band-defining and thermal filters and for the HWP), we collect about 23 modes for each 90 GHz horn, boosting the sensitivity by a factor >4 with respect to the diffraction-limited case. Given the limited aperture of the optical system, we have decided to build a cryogenically cooled telescope, to reduce the background on the detectors produced by the optical system (which is severe for multi-moded optics). The system is an on-axis refractive telescope.

Also, we have decided to fly the instrument as a stratospheric balloon payload, to significantly reduce the atmospheric background and its fluctuations. The flight will be performed during the polar night, using the earth as a giant sun-shield, and allowing the coverage of a significant fraction of the sky.

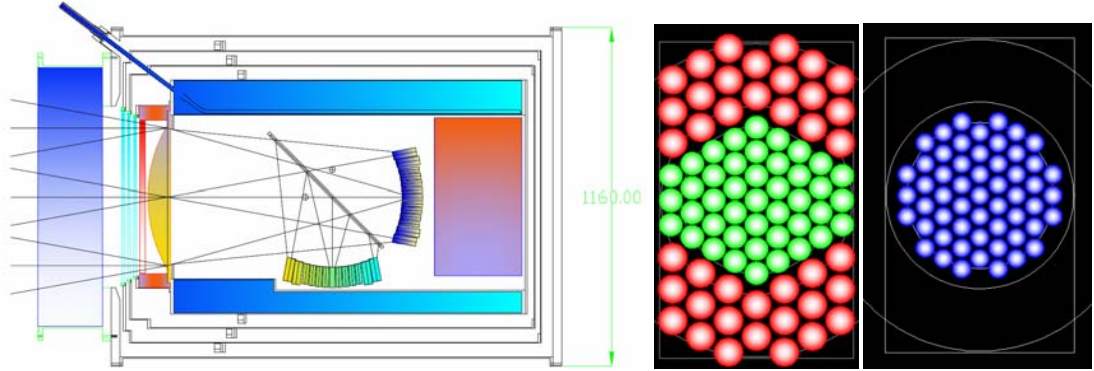

Figure 2. **Left:** Sketch of the SWIPE instrument, fully enclosed in a large liquid $^4$He cryostat. From the left are visible the large foam window (60 cm clear aperture), the three thermal filters connected to the 130K, 30K and 2K stages of the cryostat, the rotating HWP, the HDPE lens, the beamsplitting wire-grid and the two focal plane arrays of Winston horns and detectors. The rectangle on the right encloses the sub-K refrigerator. The quote is in mm. The available volume for liquid $^4$He is more than 250 liters, enough for more than 30 days of operation. The weight of the system is around 300 kg. **Center:** Layout of one of the focal planes, showing the size and location of the different detectors projected on the sky. The 43 smaller detectors are the 145GHz ones, while the 40 larger detectors are the 95GHz ones. This focal plane covers $12^\circ$ x $20^\circ$ in the sky. **Right:** same as center, for one of the two 245 GHz focal planes, with 55 detectors. This covers approximately $11^\circ$ x $11^\circ$ in the sky. The scan direction is vertical in both focal-plane plots.

We can estimate the sensitivity of such an experiment assuming photon-noise limited performance. We have studied three bands which have been used successfully at balloon altitude: (95, 145, 245) GHz, with bandwidths of (30, 45, 90) GHz respectively. We considered a configuration with a number of coupled modes, estimated by detailed electromagnetic simulations, = (23, 35, 64), and an angular resolution of (1.84, 1.48, 1.23)$^\circ$ FWHM respectively. Assuming 40% total optical efficiency, we have computed the radiative background on each detector, from the CMB itself, the telescope and the residual atmosphere at 37 km of altitude. The result is (23, 41, 102) pW, and, assuming photon-noise limited conditions, the NET is (17, 17, 20) $\mu K_{CMB}/Hz^{1/2}$. With this radiative background, photon noise limited performance can be achieved cooling the bolometers at 0.3K, avoiding the complication of 0.1K coolers.

The usable size of the focal plane and the number of detectors which can be allocated depend on issues related to the scan strategy, the needed circularity of the beams, and the acceptable levels of aberration and beam asymmetry. Moreover, accommodating in the same focal plane the three bands above would require the use of a common HWP which should operate between 80 and 290 GHz. This latter requirement is very difficult to satisfy. So we have decided to build a polarimeter for the 95 and 145 GHz bands, sharing the same focal plane, and another one for the 245 GHz band. The layout of these polarimeters is reported in Figure 2. In a multi-mode system the size of the detector apertures in the focal plane is large with respect to the size of the Airy disk. So we can use a very large focal plane, with marginal detectors up to $\pm 10^o$ off-axis. This allows us to accommodate 40 detectors at 95 GHz and 43 detectors at 145 GHz in the same focal plane. The corrected focal plane area at 245 GHz is smaller, but still allows to accommodate 55 detectors at 245 GHz. The best way to insert a polarizer in front of the detectors of the Stokes polarimeter is to mount a wire-grid with its plane tilted at 45$^o$ with respect to the optical axis, so that two focal planes are available, one transmitted and one reflected by the wire-grid. This doubles the number of detectors in each band, as evident from Figure 2. The detectors are placed in the focal plane in such a way that during each scan the same pixel of the sky is measured in both bands. 145 GHz detectors, which have a smaller aperture area, will be placed in the central region of the focal plane, to take advantage of the lower aberrations, while 95 GHz detectors will occupy the outer regions of the focal plane. A possible configuration is sketched in Fig.2 (center). The survey sensitivity for this polarimeter, assuming 25% sky coverage in 13 days of observations, is 12 μK arcmin . In terms of expected sensitivity to the power spectrum of CMB polarization, the situation is described in Fig.3. It is evident that the focus here is on the reionization bump and the rising part of the horizon-scale peak, in the region where the primordial B-modes are less contaminated by lensing B-modes.

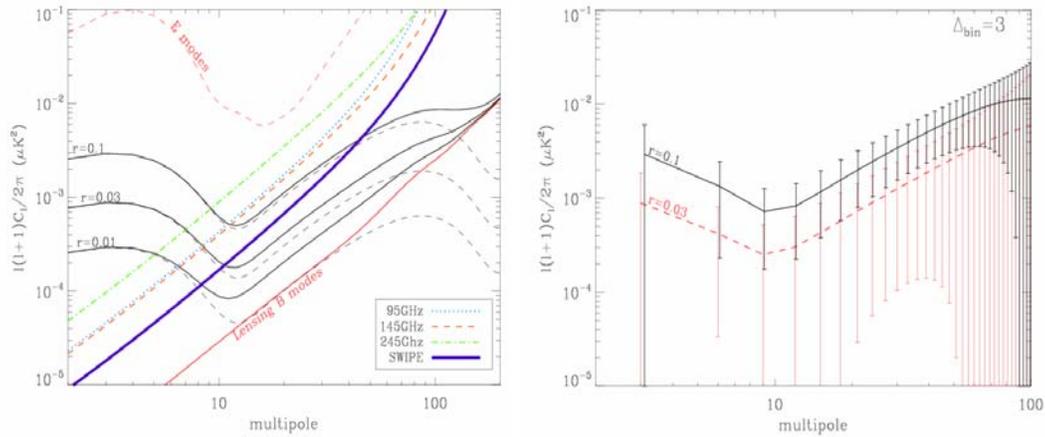

Figure 3. **Left:** Sensitivity forecast for the SWIPE survey (thick continuous line) compared to the power spectra of B-modes (thin continuous lines) for values of the tensor to scalar ratio r ranging between 0.01 and 0.1. **Right:** Forecast B-modes measurement errors, including cosmic variance.

The best way to assess the performance of this experiment is to analyze the constraints on the cosmological parameters directly from the analysis of the maps (see section 7 below).

## 4. KEY SUBSYSTEMS

### 4.1 Telescope

The optical system has been optimized to achieve high instrumental polarization purity, and to be cooled cryogenically to reduce its emission and to exploit the advantages of a cold stop. An axially symmetric refractive telescope fulfils our requirements with a low cross-polarization (< 0.2%) and a controlled instrumental polarization (the absorption polarization is less than 0.2% and the emitted component is stabilized by the use of a cold telescope). The cold stop is efficiently matched to the multi-mode patterns of the horns. This performance is maintained up to the edge of the FOV ($\pm 10°$ off-axis) for all the bands, while the constraint on the Strehl ratio (>90%) limits the focal plane radius for each band. The optics consists in a single 470 mm diameter lens and a 440 mm diameter cold stop, resulting in and entrance pupil of 450 mm diameter and $f/\# = f/1.88$. The FOV is split in polarization into two curved focal planes, both 300 mm in diameter, by a 500 mm in diameter wire grid tilted at 45$^o$. The aplanatic wide-angle lens is a simple plano-convex solution. Due to the wide spectral range covered by SWIPE, we were forced to avoid a zoned lens, which is significantly chromatic. The downside of our lens design is a maximum thickness of 62 mm, for a construction in High Density Polyethilene. This material was selected because it is easy to machine, and produces low losses in the SWIPE bands. The reflection loss, which would be of the order of 8% for an uncoated lens, is reduced by means of layers of porous PTFE used as antireflection coatings[24].

**4.2 Polarization Modulator**

The response to polarization of the whole focal plane is modulated by a large-diameter (50 cm) rotating Mesh Half-Wave Plate[25,26], which is the first device in the optical chain. These devices are based on the photolithographic technology successfully used in the past to build mesh filters and more recently mesh circular-polarizers (QWPs) at millimetre wavelengths[27,28]. These metamaterial waveplates are based on anisotropic metallic grids behaving differently along two orthogonal axes. Cascading many grids with specific geometries it is possible to produce a flat differential phase-shift of 180° across wide bandwidths. LSPE will have two mesh HWPs, one will work simultaneously across the 95 and 145 GHz frequency bands, the other will be optimized for the 245 GHz band. A 20 cm diameter prototype of a dielectrically embedded mesh-HWP working around 90 GHz is shown in Fig.4. Across a 30% bandwidth the measured cross-polarisation resulted to be below -25dB[29,30].

The HWP is rotated in a step-and-integrate mode, using a mechanical rotator similar to the one we have developed for the PILOT instrument[23]. A step motor, outside the cryostat, transmits the rotation to the phase retarder through a set of thermally insulating shafts and gears. Four pairs of optical fibers realize an absolute and reliable knowledge of the HWP position with a precision better than 0°.1.

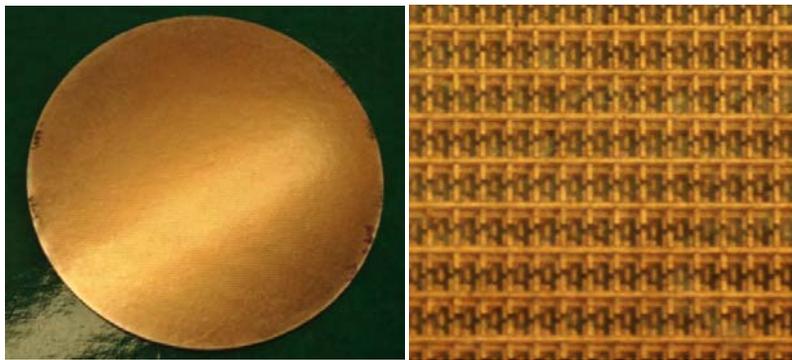

Figure 4: 20 cm diameter prototype of a mesh-HWP for the W-band (left), and a close-up of the grids (right).

**4.3 Field Optics and Detectors**

The baseline detectors for SWIPE are spider-web bolometers with Transition Edge Sensors (TESs) thermistors. Due to the very light absorber combined with the extreme electro-thermal feedback, these detectors offer reasonable time constants, despite of the large absorber area necessary to couple to a multi-mode beam. Moreover, the spider-web absorber is very efficient in rejecting primary cosmic rays present in the stratosphere and potentially very dangerous for CMB measurements[31]. A parabolic Winston horn is a classical solution for this need[32]. Unlike single-mode horns, Winston concentrators can achieve very high coupling efficiency in the main beam and, at the same time, ensure high suppression of stray radiation, through the incoherent superposition of the propagated modes at large angles. The Winston horns for SWIPE have been characterized in polarization using the HFSS software and a custom-developed polarized far field calculation code. Fig. 5 shows two orthogonal cross sections of the angular response to linearly polarized radiation. The beam ellipticity is very small, a highly desirable feature for this instrument, and will be mitigated even more coupling the horn to the telescope and its Lyot-stop. SWIPE will feature 80 horns at 95 GHz (each coupled to 23 modes) and 86 horns at 145 GHz (35 modes). Thanks to the multimode design, this setup will provide a performance comparable to that of competing instruments based on thousands of single-mode detectors.

The SWIPE arrays will use MoAu TESs, although MoCu would be suitable also, because they provide a rugged technology, which can be cycled to high temperature for outgassing, and which show no long-term ageing effects due to inter-diffusion. Achieving the needed NEP, and electrically interfacing the TESs with suitable SQUID readout technology is not the primary problem. Nor is achieving uniform, wideband matching to a multimode feed; in fact, detailed theoretical models, based on patterned absorbers coupling to multimode waveguides, and experimental demonstrations of multimode waveguide TESs are available. It should be appreciated, however, that in a multimode design, the interface with the absorber influences the beam shape and polarisation characteristics on the sky. The behaviour is intermediate between single-mode designs, where the beam characteristics are determined by the lowest-order waveguide mode, and free-space pixel designs, where the beam characteristics are determined by the geometry of the absorber. Designing multimode bolometers, and associated light pipes, is more complicated, but customised design tools are available and well established. The absorber will comprise a thin film of β-phase Ta on a suspended patterned SiN mesh to reduce heat capacity, and maintain speed. The use of MoAu TES also has the advantage that they are self passivating, and so do not require a top layer of $SiO_2$, which helps avoid long time constants. The TES wafer will be embedded in a micromachined carrier wafer, which provides a near-in electrical backshort to maintain clean optical matching across the band. The primary problems are mechanical, because each TES is physically separated from every other TES, and so it is not possible to use a single wafer that extends the whole way across each array. This problem is particularly acute at the longest wavelengths. The physical size of the array also changes the SQUID multiplexer strategy, because there must be a superconducting electrical connection between each TES and the first stage of SQUID amplification. The fact that multimode horns are used means that fewer pixels are required, and so an approach based on individual TESs and SQUIDs behind each horn becomes feasible.

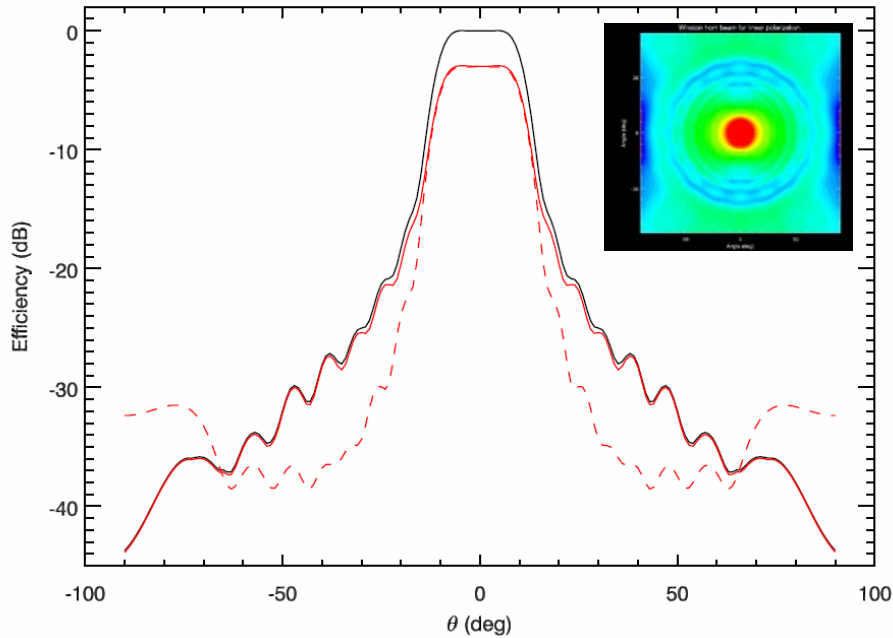

Figure 5. Far field calculation of the angular efficiency of a SWIPE Winston horn coupling 17 modes at 145GHz. The lower solid and dashed lines represent two orthogonal cross-sections for linearly polarized photons. For reference, the solid top line is the beam for unpolarized radiation. In the inset, 2D map of the same beam.

### 4.4 Cryostat

A large aluminum cryostat is used to cool down the telescope and the detectors. The system is similar to the ones we have developed for previous balloon missions[33]. The toroidal $^4$He tank is supported by a stiff structure of fiberglass tubes, and surrounded by two vapor cooled shields. The LHe tank volume is 250 liters, and detailed simulations show that the heat load on LHe can be limited to 95 mW, resulting in more than 40 days of operation. A $^3$He sorption fridge from Chase Cryogenics Inc. is used to cool-down the detectors arrays at 0.3K. A lower temperature for the bolometers is not required, given the significant radiative loading. The large vacuum window is composed by a thin (50 μm) polypropylene film, supported by a thick (20 cm) foam layer. The large optical load from the window is reflected away efficiently by a sequence of 3 mesh filters. The weight of the cryostat system, including the cold instrument, is 330 kg. The two technological challenges in this system are the size of the optical window (requiring very large filters to limit the radiative load) and the need to tilt the cryostat axis down to 70º from the vertical during observations of planets. A custom measurement campaign has started to optimize the thermal properties of the blocking / reflective filters. The cryostat tank is suspended by an array of fiberglass straps, demonstrating excellent stiffness while keeping the conduction heat load very small.

**4.5 Readout and data storage**

TES' are coupled to SQUID amplifiers. We choose to avoid multiplexing, simplifying the read-out system and reducing the read-out noise. The room temperature read-out electronics has several duties: a) TES' need to be voltage biased in order to put them into a strong electro-thermal feedback; b) SQUIDs response is highly non-linear and requires to be current biased at the critical current, and to be controlled with the flux lock loop (FLL); c) scientific data is filtered, down-sampled and sent to the acquisition system. During observations: i) when the optical load on the TES changes by more than 10% of the saturation power (typically once per day), automatic tuning and re-biasing procedure have to be re-run in order to move the operating point back to the middle of the normal-to-superconducting transition[34]; ii) every time the magnetic or the thermal conditions sensibly change (typically once per day), an automatic SQUID tuning procedure has to be run. Read-out electronics based on FPGAs are being studied[35]. The Data Acquisition System will simultaneously sample 400 differential analog inputs at a frequency of 200 Hz. Scientific, ACS and housekeeping information will be stored on the on-board SSD as 16-bit words, and a compressed sub-sample will be sent to the ground. We expect about 100 GB of uncompressed data for the whole flight. The data acquisition system is based on UEIPAC 600 units equipped with DNR-AI-225 boards. The unit is connected through a TCP socket on a RJ45 cable to a PC104 VDX-6354-PLUS PC/104+ Vortex86DX 800MHz.

## 5. MITIGATION OF SYSTEMATIC EFFECTS

|  | **Systematic effect** | **Mitigation** |
| --- | --- | --- |
| Polarization | HWP emission | Scan with HWP steady, Low temperature HWP |
|  | Wire grid emission, reflected by HWP | Low temperature WG, antireflection coating on HWP |
|  | Differential transmission of HWP | Scan with HWP steady |
|  | Differential reflection of HWP | Scan with HWP steady, antireflection coating on HWP |
|  | Differential phase shift by HWP | Scan with HWP steady, Spectral bandwidth optimization |
|  | Slant incidence of rays on HWP | Scan with HWP steady |
|  | Cross polar leakage | Lab. Calibration |
|  | Absolute polar angles calibration | Lab Calibration / Moon / Crab |
|  | Thermal fluctuations of HWP | Scan with HWP steady, thermal link HWP – cryogen |
| Optics | Main beam uncertainty | Laboratory calibration / observation of planets |
|  | Main beam ellipticity | Reduced in multimode system; lab. and flight calibration |
|  | Sidelobes pickup of sky signal | Large shields, cold stop |
|  | Sidelobes pickup of Earth and Balloon | Large shields, cold stop |
| Pointing | Pointing error | ACS Sensors |
|  | Pendulation and atmospheric emission | Not polarized / orthogonal detectors |
| Detectors | Gain uncertainty | Calibration on anisotropy |
|  | Gain stability | Calibration on anisotropy |
|  | 1/f noise | AC bias / T stabilization |
|  | Correlated thermal drift | TES |
|  | Non linearities | Compensation on scans |

Table 1 : List of potential systematic effects affecting the SWIPE instrument, and mitigation strategies.

The control of systematic effects is one of the most critical aspects of any CMB polarization experiment. The SWIPE strategy is based on solving a map of the T, Q and U Stokes parameters using each detector independently. This approach mitigates the requirements in terms of instrumental calibration, with respect to the instruments which combine data from different detectors. Still, an accurate characterization of instrumental properties is mandatory, in order to have reduce the systematic errors below the level of statistical noise. A complete analysis, based on an instrument simulator, is currently ongoing. It will cover the list of potential systematic effects reported in Table 1, including the non-trivial interaction with Galactic foreground emission. Such analysis will be used to set all the requirements in detail.

## 6. OBSERVATION AND CALIBRATION PLANS

LSPE will observe the sky by spinning at a constant speed of 3 rpm around the zenith axis. The latitude is assumed to be 78°N, as for flights launched from the Svalbard archipelago. For SWIPE, the telescope bore-sight will be at a nominal elevation of 40° and can be moved in the range from 20° to 60°, mainly to observe planets for beam calibration and investigate the effect of ground pickup. During the survey, the elevation will change every several hours in a limited range (from 30° to 40°). This, combined with the azimuthal spin, and with Earth rotation, results in a 35% coverage of the sky. Removing the regions inside the WMAP polarization mask, the SWIPE baseline coverage is 25% of the sky, which is suitable for estimation low-ell multipoles. Figure 5 shows a map of the SWIPE coverage, with the WMAP polarization mask applied. For a polarimetric survey the coverage of polarization angles is also very important. The merit function to be minimized for optimal coverage is

$$S = \sum_{pixel} W_{pixel} \qquad W_{pixel} = \frac{1}{N_{obs}^2}\left[\left(\sum_{i=1}^{N_{obs}} \cos 2\alpha_i\right)^2 + \left(\sum_{i=1}^{N_{obs}} \sin 2\alpha_i\right)^2\right] \qquad (1)$$

Where $\alpha_i$ are the observed polarization angles for each pixel and $N_{obs}$ is the number of observations for that pixel. The distribution of the operands $W_{pixel}$ for our scan strategy is shown in figure 6 (map and histogram).

Polarimeter calibration is a critical aspect for an experiment aiming at detecting the B-modes. The basic strategy for SWIPE in flight calibration is based on: (i) absolute calibration of each detector on Planck CMB anisotropies; (ii) Point Spread Function (beam) and time response estimation based on scans over planets, with different scan speed (this may require dedicated elevation changes); (iii) polarimeters calibration in terms of polarization efficiency and polarization angle, by observation of the Moon, assuming a model of the Moon polarization convolved with the instrument beam. The in-flight polarimeter calibration is used to confirm a more complete ground-based calibration, obtained illuminating the telescope by means of a polarized source in the far field.

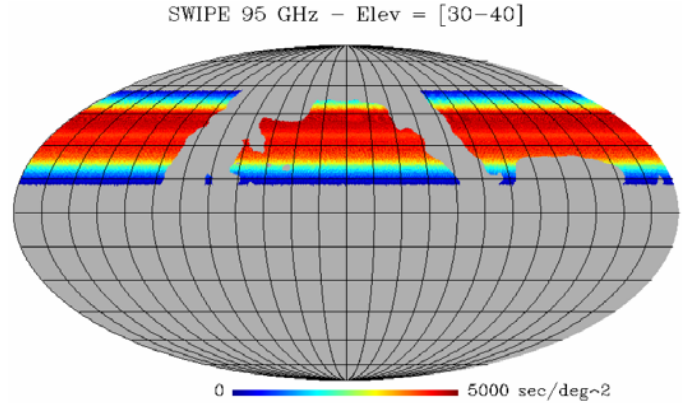

Figure 5: Sky coverage of the 95 GHz channel of SWIPE according to the baseline scanning strategy, with elevation ranging from 30° to 40°, 13 days of observation, and a flight at 78 °N latitude.

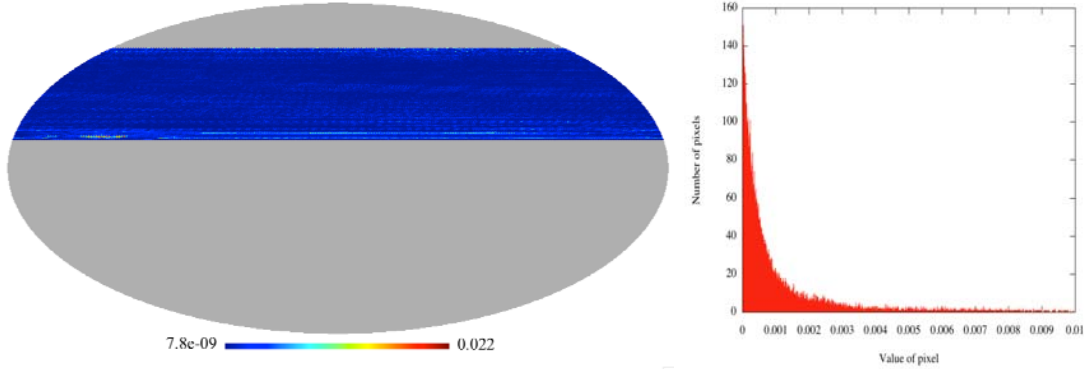

Figure 6: Distribution of the operands $W_{pixel}$ (defined in equation 1) describing the uniformity of sampling of polarization angles, for the 95 GHz channel of SWIPE, according to the baseline scanning strategy, with elevation ranging from 30° to 40°, 13 days of observation, and a flight at 78 °N latitude. **Left:** Distribution in the sky of $W_{pixel}$. **Right:** Histogram of $W_{pixel}$. The values are <<1, confirming the good uniformity of the sampled polarization angles over the whole surveyed area.

## 7. CONSTRAINING COSMOLOGICAL PARAMETERS WITH SWIPE

In this section we show the forecasts for the constraints on the tensor-to-scalar ratio $r$ and on the optical depth to re-ionization $\tau$ from SWIPE. The method we adopted to forecast the errors on the cosmological parameters is based on the publicly available Markov Chain Monte Carlo package cosmomc[36] as usual in this field. We computed the Likelihood on a 3-parameters space (the amplitude of scalar perturbations, $A_s$, the optical depth to re-ionization, $\tau$ and the tensor-to-scalar ratio $r$) fixing the remaining parameters, using a pixel-based approach described in[37]. This method is based on the direct evaluation of the likelihood in the pixel space without any power spectrum estimation step, assuming the CMB sky to be gaussian distributed. Moreover we assume a perfect foreground cleaning with negligible residuals and white noise. We leave to a future publication a deeper analysis of the effects of colored noise and realistic foreground cleaning on the cosmological parameters constrains. At each step of the Montecarlo we generate a CMB power spectrum, using publicly available CAMB Boltzman code[38], we compute the signal covariance matrix $S$ making use of the Legendre polynomials, and we evaluate the likelihood as:

$$L(m|C_\ell)\,dm = \frac{\exp\left[-\frac{1}{2}m^t(S+N)^{-1}m\right]}{|S+N|^{1/2}} \frac{dm}{(2\pi)^{3n_p/2}} \qquad (2)$$

where $m$ is the observed map ($n_p$ pixels) and $N$ is the noise covariance matrix. This approach involves the evaluation and in particular the inversion of the matrix $S+N$ that can be computationally heavy at high resolution. So we decided to perform the analysis smoothing the simulated maps at the common resolution of 7.3 degrees working at $n_{side}=16$ resolution in the HEALPix pixelization scheme[39] since the larger part of the information on $\tau$ and $r$ is contained in the lowest multipoles of the CMB power spectrum (see Figure 3). We simulated, analyzed maps and compared the results of two experiments: Spider[40], considering 4 focal plane units (FPU) at 90 and 150 GHz, 1 flight and a sky coverage of 11%, and SWIPE at 95 and 145 GHz and with a sky coverage of 22%. The results (Figure 7) show that SWIPE is able to detect $r = 0.12$ with 1-σ error of 0.032 and $\tau=0.0850$ with 1-σ error of 0.0044. This is very similar to the performance of Spider. Since the signal we are looking for is so tiny, we consider very valuable to have independent experiments, with different detection techniques measuring the sky with comparable efficiency. This is the only way to check for subtle systematic effects which, we believe, will represent the real limit of these measurements.

## 8. CONCLUSIONS

The development of SWIPE is currently approaching the end of phase-A. Focusing on large angular scales, the project promises to provide very valuable constraints on the B-modes of CMB polarization with an affordable effort, and is certainly complementary (and orthogonal) to other planned CMB polarization missions. We look forward to a flight of LSPE in the winter 2014-2015.

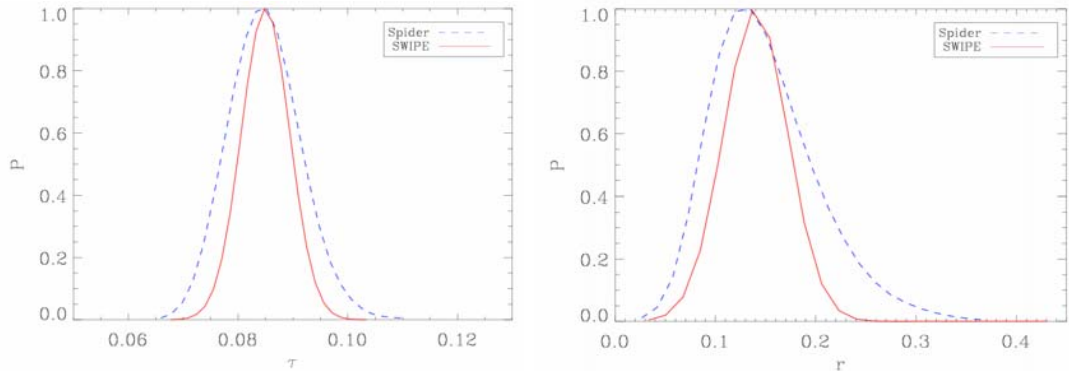

Figure 7: **Left:** Likelihood for the optical depth to reionization as estimated from Spider (dashed line) and SWIPE (continuous line). **Right:** Same for the tensor to scalar ratio. The two experiments have similar expected performance, but quite different measurement techniques.

## 9. ACKNOWLEDGMENTS

We gratefully acknowledge support from the Italian Space Agency through contract I-022-11-0 "LSPE".